\documentclass[aps,pre,10pt,twocolumn,superscriptaddress,balancelastpage,showpacs,letterpaper]{revtex4-2}
\usepackage[utf8]{inputenc}
\usepackage[T1]{fontenc}
\usepackage{amsmath,amssymb}
\usepackage{graphicx}
\usepackage{bm}
\usepackage{braket}
\usepackage{xcolor}
\usepackage{csquotes}
\usepackage{hyperref}
\hypersetup{colorlinks=true, linkcolor=purple, citecolor=teal}

\newcommand{\beq}{\begin{equation}}
\newcommand{\eeq}{\end{equation}}

\makeatletter
\let\@orig@make@capt@title\@make@capt@title
\def\@make@capt@title#1#2{\@orig@make@capt@title{{\bf #1}}{#2}}
\makeatother

\begin{document}
\title{Nonreciprocal Ising model}
\author{Yael Avni}
\affiliation{University of Chicago, James Franck Institute, 929 E 57th Street, Chicago, IL 60637}

\author{Michel Fruchart}
\affiliation{Gulliver, ESPCI Paris, Université PSL, CNRS, 75005 Paris, France}

\author{David Martin}
\affiliation{University of Chicago, Kadanoff Center for Theoretical Physics and Enrico Fermi Institute, 933 E 56th St, Chicago, IL 60637}

\author{Daniel Seara}
\affiliation{University of Chicago, James Franck Institute, 929 E 57th Street, Chicago, IL 60637}

\author{Vincenzo Vitelli}
\affiliation{University of Chicago, James Franck Institute, 929 E 57th Street, Chicago, IL 60637}
\affiliation{University of Chicago, Kadanoff Center for Theoretical Physics, 933 E 56th St, Chicago, IL 60637}

\begin{abstract}
    Systems with nonreciprocal interactions generically display time-dependent states.
    These are routinely observed in finite systems, from neuroscience to active matter, in which globally ordered oscillations exist.
    However, the stability of these uniform nonreciprocal phases in noisy spatially-extended systems, their fate in the thermodynamic limit, and the critical behavior of the corresponding phase transitions are not fully understood.
    Here, we address these questions by introducing a nonreciprocal generalization of the Ising model and study its phase transitions by means of numerical and analytical approaches. 
    While the mean-field equations predict three stable homogeneous phases (disordered, ordered and a time-dependent swap phase), our large scale numerical simulations reveal a more complex picture. Static order is destroyed in any finite dimension due to the growth of rare droplets unless the symmetry between the two spin types is broken triggering a stabilizing droplet-capture mechanism.
    The swap phase is destroyed by fluctuations in two dimensions through the proliferation of spiral defects, but stabilized in three dimensions where nonreciprocity changes the critical exponents from Ising to XY, thus giving rise to a robust spatially-distributed clock.
\end{abstract}
\maketitle
\begin{figure*}[ht]
\centering
{\includegraphics[width=0.98\textwidth,draft=false]{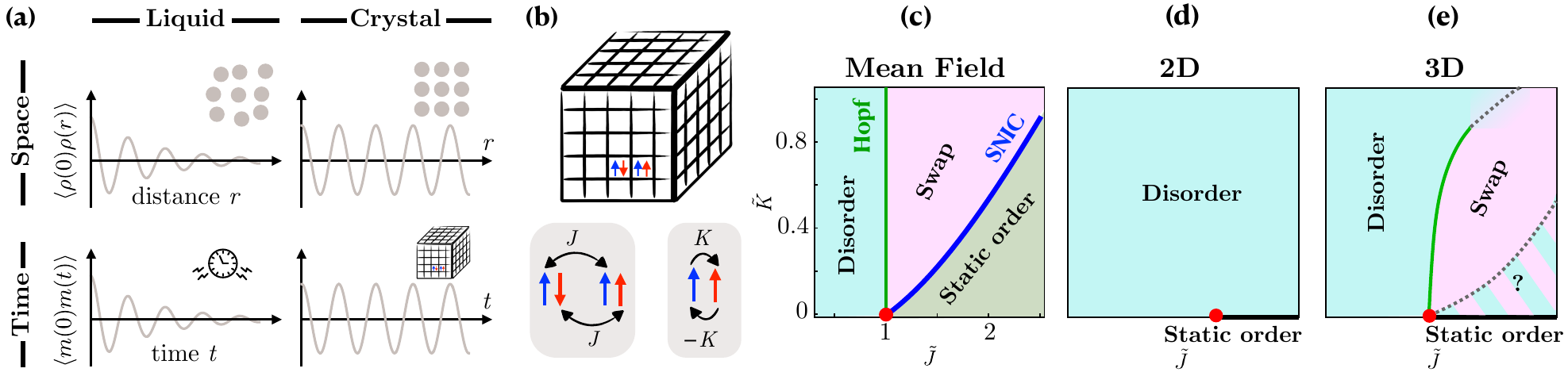}}
\caption{
\textbf{Nonreciprocal Ising model.}
(a) Comparison of noisy (left) and robust (right) clocks with different types of spatial order.
(b) Schematic drawing of the nonreciprocal Ising model. The model includes two species per site with two interaction types: 1) intra-species reciprocal nearest-neighbors interaction of strength $J$ and 2) inter-species nonreciprocal on-site interaction of strength $K$. (c) Mean-field phase diagram, (d) schematic 2D phase diagram and (e) schematic 3D phase diagram. The three phases shown on the diagrams are (i) disorder, (ii) swap and (iii) static order. The lines separating the phases are Hopf bifurcation (thin green), SNIC bifurcation (thick blue) and two yet undetermined transition lines (dashed grey). Red dot is the Pitchfork bifurcation.}
\label{Fig1}
\end{figure*}

Nonreciprocal interactions naturally arise in out-of-equilibrium systems~\cite{Meredith2020,
Ivlev2015,
Cavagna2014,
Dadhichi2020,
Guazzelli2011,
Petroff2015,
Beatus2006,
Peterson2019,
Helbing1995,
Uchida2010,
Nagy2010,
Yifat2018,
Drescher2009,
Lafferty2015,fruchart2021non,saha2020scalar, you2020nonreciprocity,frohoff2023nonreciprocal,liu2023non,brauns2023non,Loos2020,Fruchart2023} ranging from neuroscience~\cite{sompolinsky1986temporal,Derrida1987,Parisi1986} and social networks~\cite{hong2011kuramoto,hong2011conformists} to ecology~\cite{ros2023generalized,Bascompte2006,Loreau2013} and open quantum systems~\cite{Chiacchio2023,Metelmann2015,Clerk2022}.
A generic feature of these systems is the emergence of many-body limit cycles -- time-dependent states arising from non-mutual interactions between constituents. 
Such states are routinely observed in experiments and simulations of finite systems. 
A single limit cycle oscillator subject to noise eventually forgets the initial phase~\cite{delJunco2020,delJunco2020b,Cao2015,Barato2017,Fei2018,Wierenga2018,Nguyen2018,Marsland2019,Shiraishi2023,Ohga2023,Oberreiter2022}: its temporal correlations decay with time very much like density fluctuations decay with spatial separation in a liquid (Fig.~\ref{Fig1}a, left). 
In many-body systems, however, coherent oscillations may be restored over arbitrarily long periods~\cite{acebron2005kuramoto,bennett1990stability,grinstein1993temporally,Chate1995,Brunnet1994,Gallas1992,Hemmingsson1993,Grinstein1988,Bohr1987,Binder1992,Lemaitre1996,Chate1991,Chate1992,Gallas1992,Chate1997,Losson1994,Brunnet1994,Grinstein1994,Wendykier2011,Binder1997,wood2006universality,wood2006critical}: 
temporal correlations persist (without any periodic drive externally imposing a phase) very much like spatial correlations do in a crystal (Fig.~\ref{Fig1}a, right).
In this Letter, we ask: Can many-body limit cycles survive fluctuations in spatially-extended, locally-coupled nonreciprocal systems of arbitrary size? If so, what are the critical exponents of the resulting phase transitions? We address these questions by introducing a nonreciprocal generalization of what is perhaps the most paradigmatic statistical mechanical system: the Ising model.

{\it Dynamics without global optimization---}
In equilibrium, the dynamics of Ising spins 
is described by kinetic Ising models in which spins $\sigma_n=\pm1$ tend to minimize a global potential, and hence interact reciprocally~\cite{glauber1963time,walter2015introduction}.
To account for nonreciprocal interactions, we instead assume that each spin $\sigma_n$ tends to minimize its own {\it selfish energy} $E_n$, and choose the probability of flipping $\sigma_n$ at each time step to be given by the Glauber rule
\begin{equation}
    \label{FlippingProbability}
    p(\sigma_n \to - \sigma_n)
    =
    \frac{1}{2}\left[1-\tanh\left(\Delta E_n/(2k_B T)\right)\right]
\end{equation}
where $\Delta E_n$ is the change in selfish energy between both configurations, $T$ is the temperature, and $k_B$ the Boltzmann constant.
Since we define different potential functions for each spin, the dynamics cannot be derived from a single potential function and detailed balance is broken. This formulation encompasses several nonreciprocal dynamics used among the physical sciences~\cite{lynn2021broken,Mello2003,han2023controlled,Loos2023,guislain2023nonequilibrium,guislain2023discontinuous,Lima2006,Sanchez2002,Lipowski2015,garnier2024unlearnable}.

{\it The nonreciprocal Ising model---}
Systems with nonreciprocal couplings do not always exhibit macroscopic oscillations. These tend to occur, when a small number of species have competing goals.
Correspondingly, we split the spins $\sigma_n \equiv \sigma_i^\alpha$ into two species, labeled by Greek indices $\alpha=A,B$, located on sites $i$ of a $d$-dimensional cubic lattice of linear size $L$ (Fig.~\ref{Fig1}b).
Spins of the same species tend to align with their neighbors, while spins of different species interact in a nonreciprocal manner: spins $A$ tend to align with spins $B$, whereas spins $B$ tend to anti-align with spins $A$. 
This is captured by the selfish energy 
\begin{equation}
    \label{SelfishEnergy}
    E^{\alpha}_i = -J\sum_{j\,{\rm nn\,of}\,i} \sigma_i^{\alpha}\sigma_j^{\alpha} - K \varepsilon_{\alpha\beta} \sigma_i^\alpha\sigma_i^\beta
\end{equation}
with $J,K>0$, where the first term captures the \textit{intra}-species interactions with the sum running over nearest neighbors of $i$, and the second term captures the \textit{inter}-species interactions where summation over species indices $\beta$ is implied ($\varepsilon_{\alpha \beta}$ is the Levi-Civita symbol). We have assumed for simplicity that inter-species interactions only occur on-site and are purely nonreciprocal (general asymmetric on-site interactions are considered in \cite{Avni_Extended}).

{\it Mean-field equation---}
Within the mean-field approximation and upon rescaling of time and space, the dynamics of the average magnetization $m_\alpha({\vec r},t)$ is \cite{Avni_Extended}
\begin{equation}
    \label{mean_field}
    \partial_t m_{\alpha}=-m_{\alpha}+\tanh\left[\tilde{J} m_{\alpha}+\tilde{K}\varepsilon_{\alpha\beta}m_{\beta} + D \nabla^{2} m_{\alpha} \right]
\end{equation}
where $D \equiv J / (k_BT)$, $\tilde{K}=K/(k_B T)$ and $\tilde{J} \equiv 2 d J/(k_B T)$. 

The phase diagram in Fig.~\ref{Fig1}c shows the stable homogeneous solutions of Eq.~\eqref{mean_field} as a function of the couplings $\tilde{J}$ and $\tilde{K}$. When nonreciprocal interactions are turned off ($\tilde{K}=0$), a pitchfork bifurcation at $\tilde{J}_c \equiv 1$ (red point in Fig.~\ref{Fig1}c) separates a disordered phase (in blue) from a ferromagnetic phase (in green), as in the equilibrium Ising model. 
When nonreciprocal interactions are present ($\tilde{K}\neq 0$), a time-dependent oscillatory state that we dub \enquote{swap phase} arises (in pink). This limit cycle state where both species flip their magnetizations repeatedly is separated from the disordered state by a Hopf bifurcation at $\tilde{J_c}=1$ (green line), and from the ferromagnetic phase by another bifurcation known as a saddle-node on an invariant circle (SNIC) bifurcation~\cite{Strogatz2018,Izhikevich2007} (blue line).
The shape of the limit cycle in ($m_A,m_B$) space evolves from a circle near the Hopf bifurcation to a square near the SNIC bifurcation \cite{Avni_Extended}.
The oscillation period remains finite at the Hopf bifurcation, and it diverges at the SNIC bifurcation~\cite{Strogatz2018}.

{\it Monte-Carlo simulations---}
To go beyond the mean-field description, we perform large-scale Monte-Carlo (MC) simulations of the nonreciprocal Ising model in both two and three dimensions (see \cite{Avni_Extended} for details on the update rules; we did not observe qualitative differences with other update rules). Unless mentioned otherwise, we initialize the system in an ordered state where all spins of the same species are either up or down, and let the system evolve until a steady-state is reached. 

The qualitative results of our simulations are summarized in Fig.~\ref{Fig1}d-e. In 2D, any amount of nonreciprocity destroys all order. In 3D, our simulations suggest that the swap phase survives the thermodynamic limit, while the static ferromagnetic phase is eventually destroyed by any amount of nonreciprocity.

\begin{figure}[ht]
\centering
{\includegraphics[width=0.45\textwidth,draft=false]{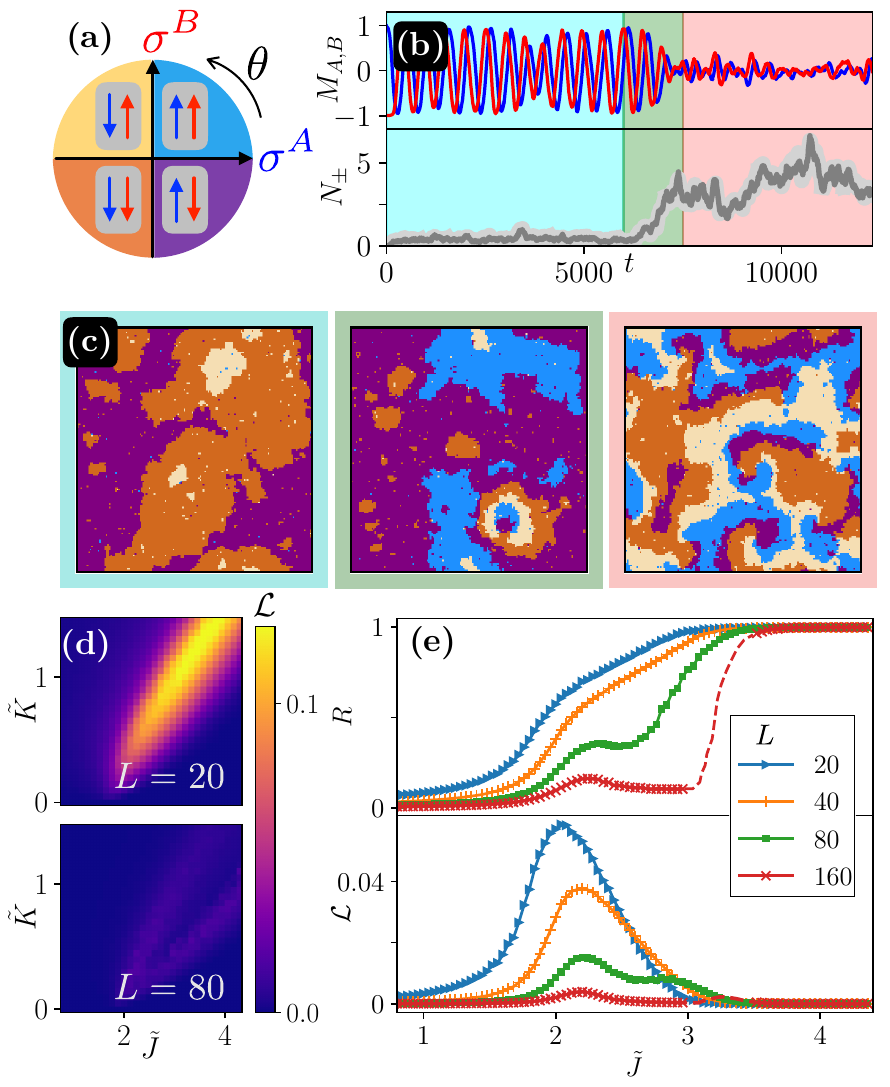}}
\caption{
\textbf{Destruction of the swap phase by spiral defects in 2D.}
(a) Color code of the angle variable, $\theta$. (b) The magnetizations (top) of species $A$ (blue) and $B$ (red) and the number of + and - defects (bottom) $N_+$ (dark grey) and $N_-$ (bright grey) averaged over 100 time steps as a function of time. MC simulation parameters: $\tilde{J}=2.8$, $\tilde{K}=0.3$, and $L=150$. We identify defects as adjacent 2 by 2 sites with a cycle of $\uparrow \uparrow\to\uparrow \downarrow\to\downarrow \downarrow\to\downarrow\uparrow$, either clockwise ($+$ defects) or anti-clockwise ($-$ defects). (c) Snapshots of $\theta$ before (left), at the onset (middle), and after (right) the proliferation of spiral defects (frame color corresponds to colored regions in panel b). See Movie 1 for the full evolution. (d) Color map of phase space angular momentum, $\mathcal{L}$, as a function of $\tilde{J}$ and $\tilde{K}$, for linear system size $L=20$ (top) and $L=80$ (bottom). (e)
Synchronization order parameter $R$ (top) and $\mathcal{L}$ (bottom) as a function of $\tilde{J}$ for $\tilde{K}=0.3$ and different $L$. Dashed red line shows upper bound for $R$ and $\mathcal{L}$ for points that did not converge during simulation running time.}
\label{Fig2}
\end{figure}

In order to distinguish the different possible phases, we introduce the synchronization order parameter~\cite{acebron2005kuramoto}
\begin{equation} \label{R_definition}
R\equiv \left\langle s\right\rangle_{t,\Omega}
\,\,\,\,\text{with}\,\,\,
s \equiv \frac{1}{L^d}\bigg|\sum_{j}{\rm e}^{i\theta_j}\bigg|= \sqrt{\frac{M_A ^2 + M_B ^2}{2}}
\end{equation}
and the phase space angular momentum related to entropy production rate~\cite{seara2021irreversibility,Avni_Extended}
\begin{align} \label{Angular_momentum}
\mathcal{L}\equiv\langle \ell\rangle_{t,\Omega}\,\,\,{\rm with}\,\,\,\ell & \equiv M_{B}\partial_{t}M_{A}-M_{A}\partial_{t}M_{B} 
\end{align}
where $\theta_j$ is the angle on the $(\sigma_j^A,\sigma_j^B)$ plane (Fig.~\ref{Fig2}a), $M_\alpha=\sum_j \sigma_j^\alpha/L^d$ are the total magnetizations and the average $\langle ...\rangle_{t,\Omega}$ is over time and realizations.
The synchronization order parameter $R$ is zero if the system is disordered and nonzero in both the swap and static-order phases, while $\mathcal{L}$ is zero in the disordered and static-order phases and non-zero in the swap phase~\cite{Avni_Extended}.

{\it Destruction of the swap phase in 2D by spiral defects---}
In small 2D systems, simulations show states in which $M_A$ and $M_B$ oscillate in time (like in the mean-field swap phase) with a large amplitude. However, oscillations are irregular, and typically occur through nucleation of droplets of opposite magnetization by the \enquote{unsatisfied} species, in alternating order (see Movie 1). 
As the size $L$ increases, it becomes apparent that this oscillatory state is a transient, eventually destabilized by proliferation of spiral defects similar to the rotational-symmetric spirals observed in the complex Ginzburg-Landau equation~\cite{aranson2002world,aranson1998spiral,Aranson1993,grinstein1993temporally,altman2015two,wachtel2016electrodynamic,chate1996phase,tan2020topological,liu2021topological,Michaud2022,Brauns2021,zhang2023pulsating,manacorda2023pulsating}, but whose shape has 4-fold rotational symmetry (four \enquote{arms}).
This is evidenced in Fig.~\ref{Fig2}b-c, which shows the emergence of defects in the angle $\theta$ and the resulting drop in the magnetizations $M_A$ and $M_B$ after an oscillatory transient, along with snapshots of $\theta$ at different stages (see also Movie 1). 

Quantitatively, the absence of a swap phase in the thermodynamic limit can be seen from the behavior of the order parameters as system size increases. Figure.~\ref{Fig2}d shows a color map of $\mathcal{L}$ as a function of $\tilde{J}$ and $\tilde{K}$. While a regime in which $\mathcal{L}\neq 0$, corresponding to the swap phase, appear at $L=20$, it almost diminishes completely at $L=80$ . A more thorough analysis is shown in Fig.~\ref{Fig2}e where both $R$ and $\mathcal{L}$ are shown as a function of $\tilde{J}$ and fixed $\tilde{K}$ for different system sizes.
In the intermediate region corresponding to the swap phase ($1.5\lesssim\tilde{J}\lesssim 3$), both $R$ and $\mathcal{L}$ go to zero as $L$ increases, thereby indicating the destruction of the swap phase.
In addition, the putative critical $\tilde{J}$ marking the transition from disorder to swap depends on the system size, signifying the absence of a well-defined phase transition in the thermodynamic limit.

\begin{figure*}[ht]
\centering
{\includegraphics[width=1\textwidth,draft=false]{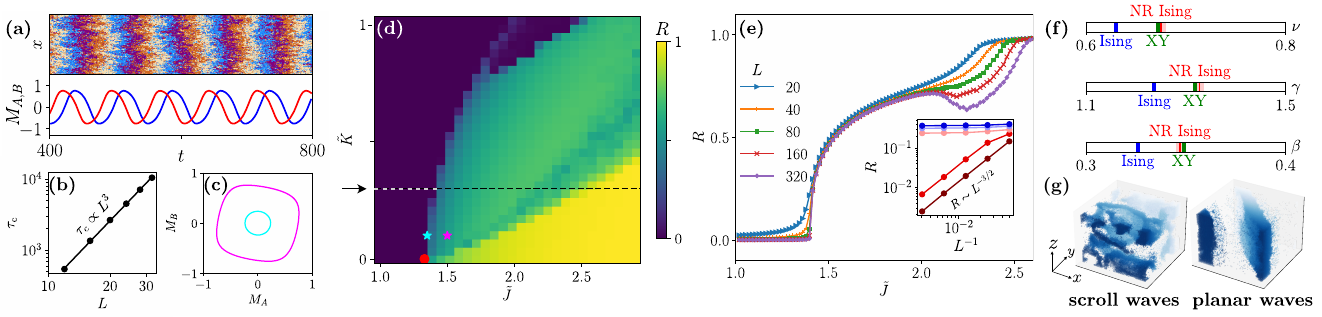}}
\caption{
\textbf{Stability of the swap phase in 3D and critical exponents.}  (a) Time evolution of a stable swap state with $\tilde{J} = 1.5$, $\tilde{K} = 0.1$ and $L=160$. Top: a kymograph of $\theta$ (color code in Fig.~\ref{Fig2}a) for a single 1D row in the 3D system. Bottom: $M_A$ (blue) and $M_B$ (red) as a function of time. (b) Coherence time $\tau_c$ as a function of system size. $\tau_c$ is defined as the time for which the envelope of the correlation function $C(\tau)\equiv \langle M_A(t)M_A(t+\tau)\rangle_t$ decreases to a factor of $e^{-1}$ of its initial value. Simulation parameters are $\tilde{J}=1.6$ and $\tilde{K}=0.1$. (c) $M_B$ as a function of $M_A$ at steady state for the points marked by stars in panel d.
(d) Color map of $R$ as a function of $\tilde{J}$ and $\tilde{K}$ for $L=320$. Red dot is the pitchfork bifurcation.  (e) $R$ as a function of $\tilde{J}$ for $\tilde{K} = 0.3$ (dashed line in panel d) shown for different linear system size $L$. 
Inset: $R$ vs. $L^{-1}$ for $\tilde{J}=1.38, 1.40, 1.42,  1.43, 1.45$ from bottom to top on a log-log scale. (f) Critical exponents of the 3D nonreciprocal Ising model (NR Ising) for $\tilde{K}=0.1$ extracted from finite-size scaling analysis along with their standard deviation represented by a semi-transparent red rectangle, and compared with the 3D Ising model and the 3D XY model. (g) Simulation snapshots of the 3D system showing scroll waves and planar waves. Sites in $\uparrow\uparrow$ state are shown in blue while other sites are not shown. System parameters: $\tilde{J}=2.28$, $\tilde{K}=0.3$, $L=80$.}
\label{Fig3}
\end{figure*}

{\it Existence of a stable swap phase in 3D and critical exponents---}
In 3D, we find that a stable swap phase does exist. 
Its behavior is demonstrated in Fig.~\ref{Fig3}a where we show a kymograph of the $\theta$-field of a single 1D row in the 3D lattice as well as plot the total magnetizations as a function of time (see also Movie 2). The oscillations are spatially homogeneous and $M_A(t)$ and $M_B(t)$ have a fixed period and phase shift. 
The coherence time of $M_A$ and $M_B$ diverges with system size as $\tau_c\propto L^d$ (Fig.~\ref{Fig3}b), in agreement with a temporal crystal behavior (Fig.~\ref{Fig1}a)~\cite{oberreiter2021stochastic}. 
In the thermodynamic limit, coherent oscillations persist indefinitely despite the presence of noise.
 
Simulations with varying system sizes reveal a well-defined phase transition between a disordered phase with $R=0$ to a swap phase with non-zero $R$. 
Figure~\ref{Fig3}d shows a color map of $R$ in the $(\tilde{J},\tilde{K})$ space, for $L=320$, while Fig.~\ref{Fig3}e shows a cut of the color map for $\tilde{K}=0.3$, for different $L$. 
Unlike in 2D (Fig.~\ref{Fig2}e), here there is a critical $\tilde{J}$, below which $R\sim L^{-d/2}$ and above which it converges to a non-zero value (Fig.~\ref{Fig3}e, inset). The stability of the swap phase in 3D holds even when the inter-species symmetry is broken by the addition of an on-site reciprocal coupling between the species, see Ref.~\cite{Avni_Extended}. 
 
Figure~\ref{Fig3}c shows that the phase-space trajectories become circular when approaching the phase transition line, supporting a Hopf behavior \cite{kuramoto1984chemical}. Renormalization group studies based on the $\epsilon$-expansion suggest that the universality class associated to the Hopf bifurcation is formally equivalent to the universality class of the ferromagnet/paramagnet transition of the XY model~\cite{risler2004universal,risler2005universal}. 
Using finite-size scaling \cite{Avni_Extended} and setting $\tilde{K}=0.1$ we determine the critical exponents of the correlation length, susceptibility, and order parameter: $\nu=0.675\pm0.005$, $\gamma=1.328\pm0.009$, and $\beta=0.347\pm0.002$, which are indeed in good agreement with 3D XY critical exponents~\cite{pelissetto2002critical,campostrini2001critical}, more so than with 3D Ising critical exponents~\cite{pelissetto2002critical} corresponding to the $\tilde{K}=0$ case, see Fig.~\ref{Fig3}f.

For a fixed $\tilde{K}$, the critical $\tilde{J}$ separating disorder from swap depends on $\tilde{K}$, in contrast with the mean-field prediction, in which $\tilde{J}_c$ is independent of $\tilde{K}$ (Fig.~\ref{Fig1}c). Above some critical ${\tilde K}$ value, the phase transition loses its second-order-like nature (see $\tilde{K}\gtrsim 0.75$ region in Fig.~\ref{Fig3}d where $R$ becomes discontinuous), apparently due to scroll waves (3D analog of spiral waves)~\cite{winfree1980geometry,Winfree1984,Winfree1983a,Winfree1983b,Winfree1983c} that destabilize the swap phase~\cite{Avni_Extended}. Within the size limitations of our simulations, it is unclear what part of the swap phase is destroyed by scroll waves in the thermodynamic limit and what type of phase transition is associated with their appearance~(Fig.~\ref{Fig1}e).
 
We note that with random initial conditions, the system can coarsen into long-lived scroll waves and planar waves (Fig.~\ref{Fig3}g), even when ordered initial conditions would otherwise lead to global oscillations (\cite{Avni_Extended} and Movie 3).

\begin{figure}[ht]
\centering
{\includegraphics[width=0.5\textwidth,draft=false]{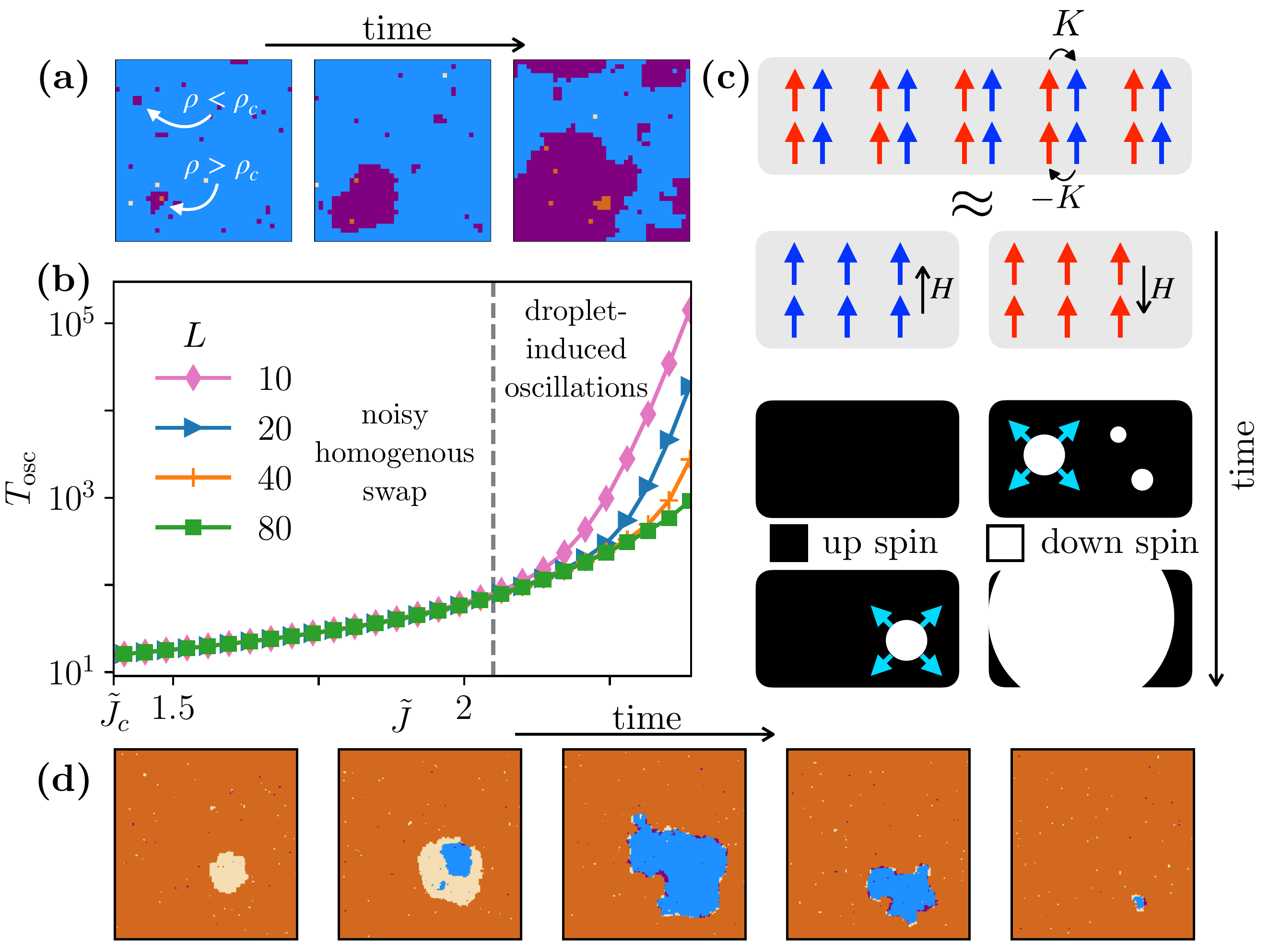}}
\caption{
\textbf{Destruction of the static-order phase due to droplet growth.} (a) Snapshots of the $\theta$ field (color code in Fig.~\ref{Fig2}a) in a 2D system, taken 20 (left) 100 (middle) and 150 (right) MC sweeps after initialization in an ordered state. The largest droplet at $t=20$ is shown to expand in later times, while the second largest droplet, which did not exceed the critical droplet size, $\rho_c$, shrunk and disappeared in later times. System parameters: $\tilde{J} = 2.8$, $\tilde{K} = 0.3$, $L = 40$. (b) Oscillation period of the total magnetizations $T_{\rm osc}$ in the 3D system as a function of $\tilde{J}$ for fixed $\tilde{K}=0.3$. $T_{\rm osc}$ is measured as twice the average time between two subsequent sign flips of $M_A$. The period is finite at the critical point $\tilde{J}_{\text{c}}$, supporting a Hopf-like behavior. Dashed grey line separates a regime in which the oscillations are driven by well-defined droplets (right) from a regime in which they do not (left). (c) Schematic drawing of the droplet argument for the instability of the static-order phase. A system in the static-order phase (first row) can be reduced into two equilibrium Ising models subject to magnetic fields (second row), where one system is in a stable equilibrium while the other in a metastable state. The metastable system nucleates droplets that expand and flip the magnetization (third row). The stable system becomes metastable and nucleates droplets that expand with the same velocity (fourth row), and so on.
 (d) Snapshots of $\theta$ in a 2D system with broken symmetry between the spin species. Simulation parameters: $\tilde{J}=3.3$, $\tilde{K}_+=0.1$, $\tilde{K}_-=0.3$, and $L=150$. See Movie 5 for the full evolution. Species $B$ nucleates a droplet, but before it reaches system size, a nested droplet of $A$-spins nucleates and expands more rapidly, eventually catching up with the boundaries of the $B$-droplet, making it shrink and disappear. This mechanism can stabilize the static-order phase when $\tilde{K}_+<\tilde{K}_-$~\cite{Avni_Extended}.
}
\label{Fig4}
\end{figure}

{\it Instability of the static-order phase due to droplet growth in any finite dimension---}
In finite systems, a ferromagnetic-like state with static order is observed both in 2D and 3D. As system size increases, however, this static state is destabilized by nucleation of droplets (see Fig.~\ref{Fig4}a and Movie 4) that grow and flip the magnetization, in alternating order of A- and B-spins. As a result, the static-order phase is replaced by droplet-induced oscillations (different in nature from the homogeneous noisy oscillations close to the Hopf bifurcation shown in Fig.~\ref{Fig3}a and Movie 2) with oscillation period $T_{\rm osc}$ that converges to a finite value as system size increases (Fig~\ref{Fig4}b). This is further supported by Figs.~\ref{Fig2}e and \ref{Fig3}e (right parts) showing that the $\tilde{J}$-transition-point between static order and oscillations in finite systems (indicated by a sharp change in the slope of $R$) increases with size.

To see why nonreciprocity allows droplets to grow, in contrast with the equilibrium Ising model where they tend to shrink, assume first that a system with small nonreciprocal coupling $K$ is in the static-order phase, so most spins in both lattices are up~\cite{assis2011infinite}. Since the system is static, it can be mapped into two equilibrium Ising models (Fig.~\ref{Fig4}c) with opposite magnetic fields, with energy $E = -J\sum_{\langle i,j\rangle} \sigma_i \sigma_j - H\sum_{i}\sigma_i$ where $H\approx K$ for A-spins and $H\approx-K$ for B-spins (see Eq.~(\ref{SelfishEnergy})). While sub-system A is in a stable state in the effective equilibrium system, sub-system B is in a metastable state, as it prefers a state with opposite magnetization. For large enough system size, sub-system $B$ then transitions to its equilibrium stable state by nucleating droplets larger than a critical size $\rho_c$ whose value depends on $\tilde{J}$ and $\tilde{K}$~\cite{rikvold1994metastable,assis2011infinite}. After a droplet has expanded beyond $\rho_c$, the stable sub-system becomes metastable and nucleates droplets, and so on (Fig.~\ref{Fig4}c). Crucially, for finite $\tilde{J}$ and $\tilde{K}$, $\rho_c$ is finite, making the static-order phase unstable in the thermodynamic limit in any finite dimension. Note that when the symmetry between the spin species is broken, droplets do not expand at the same velocity, leading to droplet-capture and shrinkage that can re-stabilize the static-order phase (Fig.~\ref{Fig4}d and Movie 5). This can be achieved, for example, by modifying the inter-species interaction in the selfish energy, Eq.~(\ref{SelfishEnergy}), to be $-K_{\alpha \beta}\sigma_i^{\alpha}\sigma_i^{\beta}$ where $K_{AB}=K_++K_-$ and $K_{BA}=K_+-K_-$ (see Ref.~\cite{Avni_Extended} for details).

What is the fate of the droplet-induced oscillations regime occurring at high $\tilde{J}/\tilde{K}$? In 2D it is unstable due to spirals (Fig.~\ref{Fig2}). In 3D, it exhibits large structures with out-of-phase regions \cite{Avni_Extended}, in contrast with the more homogeneous swap phase occurring at lower $\tilde{J}/\tilde{K}$ as in Fig.~\ref{Fig3}a. Moreover, in this regime the degree of synchronization in 3D, while clearly decreasing, does not converge with system size (Fig.~\ref{Fig3}e, $\tilde{J}\gtrsim 2$). It is therefore unclear whether the resulting phase is disordered or oscillating, but the static order can be ruled out (Fig.~\ref{Fig1}e).

We have found that the only stable phase in 2D and 3D, other than disorder, is the 3D swap phase. 
This can be traced to the fact that the swap phase spontaneously breaks a continuous symmetry (continuous time-translation invariance) rather than a discrete one~\cite{Avni_Extended}.
In this case, an analogy with the Mermin-Wagner theorem~\cite{Mermin1966,Hohenberg1967} in the time domain~\cite{chan2015limit,daviet2024nonequilibrium,lang2024field} suggests that order is destroyed in $d\leq 2$ by the fluctuations of the Goldstone modes associated with the broken continuous time translation invariance, but is possible in $d \geq 3$.
In addition, contrary to discrete symmetries, phases with spontaneously broken continuous symmetries do not support well-defined droplet excitations (domain walls are progressively blurred as time evolves) and are therefore stable against droplet growth \cite{Avni_Extended,grinstein1993temporally,Grinstein1994,benvegnen2023metastability}.

We have shown that a nonreciprocal generalization of the 3D Ising model can act as a stable spatially-distributed clock characterized by well-defined critical properties reminiscent of time-crystals~\cite{Khemani2019,Yao2020,Zaletel2023,winfree1980geometry,Sacha2017,Wilczek2012,Shapere2012,evers2023active,kongkhambut2022observation,wu2024dissipative}. While minimalistic, this model contains features arising in models of the human brain~\cite{lynn2021broken}, opinion dynamics~\cite{masuda2013voter,castellano2009statistical}, spinor BECs~\cite{Buca2019,Dogra2019,Chiacchio2023} and micromechanical oscillators~\cite{han2023controlled}, which can all be modeled by non-potential spin systems.

\medskip
\noindent \emph{Code availability} --- The code used for performing the Monte-Carlo simulations, computing the critical exponents, and plotting the mean-field phase portrait is available under the 2-clause BSD license at \url{https://doi.org/10.5281/zenodo.14816551}

\bigskip\bigskip

\begin{acknowledgments}
We thank G. Biroli, S. Diehl, O. Granek, M. Han, T. Khain, P. Littlewood, R. Mandal, D. Mukamel, S. Sethi, S. Sondhi, G. A. Weiderpass, C. Weiss, and T. Witten for helpful discussions.
Y.A., D.S. and M.F. acknowledge support from a MRSEC-funded Kadanoff–Rice fellowship and the University of Chicago Materials Research Science and Engineering Center, which is funded by the National Science Foundation under award no. DMR-2011854. Y.A. acknowledges support from the Zuckerman STEM Leadership Program. 
D.M., M.F., and V.V acknowledge support from the France Chicago center through a FACCTS grant. M.F. acknowledges support from the National Science
Foundation under grant no. DMR-2118415 and the Simons Foundation.
V.V. acknowledges partial support from the Army Research Office under grant nos. W911NF-22-2-0109 and W911NF-23-1-0212 and the Theory in Biology program of the Chan Zuckerberg Initiative. This research was partly supported by the National Science Foundation through the Center for Living Systems (grant no. 2317138) and the National Institute for   Theory and Mathematics in Biology (NITMB).
All the authors acknowledge the support of the UChicago Research Computing Center which provided the computing resources for this work.
\end{acknowledgments}

%%%%%%%%
\end{document}